\documentclass[conference]{IEEEtran}
\IEEEoverridecommandlockouts
\usepackage{cite}
\usepackage{amsmath,amssymb,amsfonts}

\usepackage{mathtools}

\usepackage{cuted}
\usepackage{graphicx}
\usepackage{textcomp}
\usepackage{xcolor}

\usepackage{amsmath, amssymb, amsthm,algorithm}
\usepackage[latin1]{inputenc}

\usepackage{latexsym}
\usepackage{graphics,epsfig}
\usepackage{amsfonts}
\usepackage{booktabs}
\usepackage{color}
\usepackage{multirow}
\usepackage[justification=centering]{caption}
\usepackage{dsfont}
\usepackage{graphicx}
\usepackage{adjustbox}
\usepackage{kantlipsum}
\usepackage[caption=false,font=footnotesize]{subfig}
\usepackage{cases}
\usepackage{url}
\usepackage{tabu}
\usepackage{makecell}
\usepackage{soul}

\usepackage{array}
\usepackage{algpseudocode}
\usepackage{mathtools,lipsum,cuted}

\usepackage[justification=centering]{caption}

\usepackage[caption=false,font=footnotesize]{subfig}
\usepackage[T1]{fontenc}
\usepackage{mwe}
\usepackage{subfig}

\newlength{\tempheight}
\newlength{\tempwidth}

\newcommand{\rowname}[1]
{\rotatebox{90}{\makebox[\tempheight][c]{#1}}}

\newcommand{\columnname}[1]
{\makebox[\tempwidth][c]{#1}}

\makeatletter
\def\BState{\State\hskip-\ALG@thistlm}
\makeatother

\newcounter{example}[section]

\usepackage[english]{babel}
\usepackage{verbatim}

\usepackage[english]{babel}

\theoremstyle{plain}

\theoremstyle{remark}

\def\BibTeX{{\rm B\kern-.05em{\sc i\kern-.025em b}\kern-.08em
    T\kern-.1667em\lower.7ex\hbox{E}\kern-.125emX}}
\begin{document}

\title{Towards Optimal Serverless Function Scaling \\in Edge Computing Network}

\author{\IEEEauthorblockN{Mounir Bensalem$^{*}$,  Francisco Carpio$^{*}$ and Admela Jukan$^{*}$}
\IEEEauthorblockA{$^{*}$Technische Universit\"at Braunschweig, Germany;
\{mounir.bensalem, f.carpio, a.jukan\}@tu-bs.de}

}

\maketitle

\begin{abstract}
Serverless computing has emerged as a new execution model which gained a lot of
attention in cloud computing thanks to the latest advances in containerization
technologies. Recently, serverless has been adopted at the edge, where it can
help overcome heterogeneity issues, constrained nature and dynamicity of edge
devices. Due to the distributed nature of edge devices, however, the
\emph{scaling} of serverless functions presents a major challenge. We address
this challenge by studying the optimality of serverless function scaling. To
this end, we propose Semi-Markov Decision Process-based (SMDP) theoretical
model, which yields optimal solutions by solving the serverless function scaling problem as a decision making problem. We compare the SMDP solution with practical,
monitoring-based heuristics.  We show that SMDP can be effectively used in edge
computing networks, and in combination with monitoring-based approaches also in
real-world implementations. 
\end{abstract}


\section{Introduction}

In the cloud, Function-as-a-Service (FaaS) computing model provides for fast
autoscaling, by removing decision-making on scaling thresholds; reduces starting
times, by using containerization as underlying technology; and provides easy
capacity planning, by only charging when resources are used; albeit at costs of losing
control on the deployment environment by the developers. In edge computing,
however, such scaling of services is challenged by the distributed nature and
heterogeneity of devices in edge network nodes as well as the resulting
performance variability. To overcome these limitations, the serverless execution model with scaling features is
envisioned as a prime mechanism in edge computing networks.

Studies adapting and further developing the well-known cloud computing scaling
mechanisms are few and far between in the edge context. Since function provider
mechanism can scale based on different parameters, such as by the number of requests
per second, CPU or memory resource utilization among others, determining which
mechanism is more appropriate in the edge is an open challenge. Considering that
edge nodes are distributed, constrained computing and networked units, heterogeneous and
volatile, optimal scaling methods are non-trivial on these type of
dynamic networked systems.

In this paper, we address the optimality of serverless scaling in edge computing network. We propose  Semi-Markov Decision Process-based (SMDP) model for the scaling problem of serverless functions as a decision making problem, with actions of scaling the functions up or
down, the reward depending on processing and queueing costs. The SMDP model maximizes the long-term expected reward of the system,
considering the system gains, costs of queueing and resource utilization.  SMDP enables us to obtain the optimal
solution where it can be also decided on the next action at each system state. From the practical
perspective, our model does not need to know which function is running in which
node, thus making the telemetry in edge network rather practical. 
The theoretical results are compared with practical, monitoring-based algorithms.
We show that this novel application of SMDP can be effectively used for function scaling, and in
combination with monitoring-based approaches also in real-world edge network implementations, e.g. with OpenFaaS.


The rest of the paper is organized as follows. Section \ref{sec:related}
describes the related work. Section \ref{sec:model} introduces the SMDP model.
Section \ref{sec:results} evaluates the performance. Section
\ref{sec:conclusion} concludes the paper.

\section{Related Work} \label{sec:related}

Function scaling, -  a process of increasing or decreasing a number of function
replicas, - is solved in a centralized manner both commercially, such as in
Amazon AWS Lambda or Google Cloud Functions, and in open source serverless
platforms, such as OpenFaaS or Apache OpenWhisk. While previous work adopts
these platforms to the edge context, including \cite{Carpio2022},
\cite{baresi2019towards},  there is lack of specific scaling mechanisms designed
to work for this type of systems. Related theoretical works studying aspects
such as job scheduling \cite{tan2017online}, resource provisioning
\cite{ascigil2021resource} and placement of resources \cite{Bensalem2020} have
been proposed in this context. These solutions, however, assume periodic
collection of telemetry data to provide optimal solutions, such as resources
state and users' demand of functions. This requires linear programing models,
which are known to be computationally demanding. From the practical perspective,
computationally complex optimization models cannot adequately consider software
limitations in terms of decision times, complexity of the related telemetry,
etc. In addition, joint solutions for functions of scaling, edge node allocation
and scheduling are complex to implement, as they use multiple and diverse
multiple tools when orchestrating services.

To the best of our knowledge, this paper is the first to study optimal scaling
serverless functions problem theoretically by using SMDP modeling. We found
inspiration to using SMDP model from a few related works, that used this method
in different context. In \cite{liang2020reinforcement} and \cite{zheng2015smdp},
SMDP-based resource allocation schemes were used in vehicular networks for
quality of experience. The paper considers the service requests from vehicles
and decides whether to process it locally or to transfer to other nodes,
considering the reward and constraints of each possible action. The reward
consists of the income and costs of power consumption and processing time. In
\cite{li2018smdp}, SMDP was used for coordinated virtual machine (VM) allocation
for a cloud-fog computing system, considering the balance between the high cost
of communication to the cloud and and the limited fog capacity. 

\begin{figure}[!t]
  \centering
  \includegraphics[width=0.9\columnwidth]{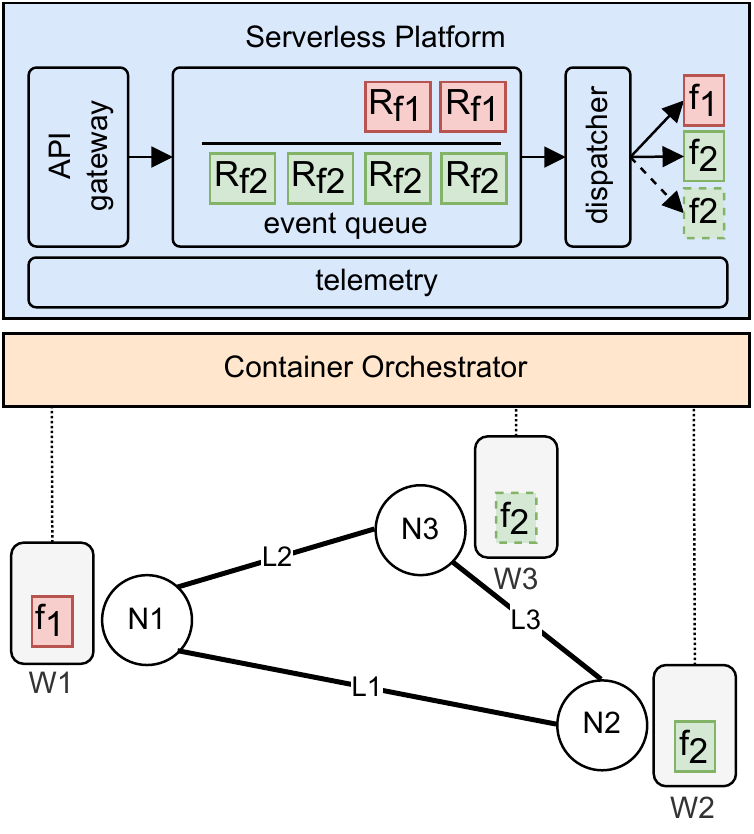}
  \caption{Serverless system at the edge}
  \label{fig:system}
\end{figure}

\section{SMDP-based Scaling model}\label{sec:model}

Our system model is illustrated in Fig. \ref{fig:system}. It consists of a set
of edge computing nodes (i.e., W1, W2 and W3),  serverless  platform and
container orchestrator (e.g., Kubernetes), all generally located in different
parts of the network. The serverless platform (e.g., OpenFaaS) is instantiated
on the cluster with access to all the resources and bundles the API gateway,
event queue, dispatcher and telemetry components. The API gateway handles HTTP
requests from the users requesting functions and creates events. The event
queues allocates different queues per type of function requests where they wait
to be served. The dispatcher is responsible for scaling decisions to be made per
type of function based on the number of requests and on the resource
utilization. For instance, in openFaaS, we monitor the arrival rate (or load)
and whenever it exceeds a threshold for a specific function, new replicas are
created; whenever the queue is empty it removes the replicas. The information
about arrival event and the queue information is collected by the telemetry
component through an API exposed by the container orchestrator. The scaling
decisions are sent to the container orchestrator which actually creates or
removes replicas of functions from the nodes, depending on whether the system is
scaling up or down. The container orchestrator is responsible for placing the
function replicas on edge nodes. 

\subsection{Problem Formulation}

We assume a single master deployment and a set of edge nodes denoted as
$\mathcal{E}=\{E_{1},..,E_{n},...,E_{N}\}$, where $E_{n}$ represents the
$n^{\text{th}}$ edge node. Each edge node has a certain amount of capacity which
we refer to as a number of CPU units $C_{n}$. Every specific function requires a
certain amount of resources. Thus, we categorize all function requests into $K$
classes based on their resource needs. A function request of class $k$, where
$1\leq k\leq K$, is denoted as $f_k$, and requires $b_k$ CPU units,  whereby
$b_k\leq max_{\forall n: n\in[1,N]}\{C_n\}$, i.e., for any function $f_k$, there
exist at least one edge node $E_{n}$ with enough capacity $C_n$ to satisfy the
request. We model the arrival and service processes of function requests of
class $k$ as a Poisson process with rates $\lambda_k$, and $\mu_k$,
respectively. We assume that all function requests are queued in a buffer with
infinite capacity. The proposed SMDP model is described by the components \{$S$,
$A(s)$, $p(s'|s,a)$, $r(s|a)$\}, where $S$ is the state space, $A(s)$ is the set
of feasible actions at the state $s\in S$, $p(s'|s,a)$ is the transition
probability from the state $s$ to the state $s'$  when an action $a$ is chosen,
and $r(s|a)$ is the reward of the system at the state $s$ when choosing the
action $a$. 

\subsection{System States}

The system state $s$ represents the number of replicas of each function $k$ in
all nodes $E_n$,  the number of function requests of each service class $k$ in
the queue, and the a set of events that can happen in the system:
\begin{equation}\label{eq:state}
\begin{split}
S=\{s|s=& (\Delta, Q, e)\},
\end{split}
\end{equation}
where a set $\Delta=\delta_{1},...,\delta_{k},...,\delta_{K}$ indicates a number
of functions, where the variable $\delta_{k}$ denotes the number of functions of
class $k$ replicated in all nodes.  $Q=\{ Q_1,...,Q_K\}$  denotes the function
request queue length vector. The variable $e$ describes an event that occurs in
the system, such as $e = \{Ar, D\}$, where a set $Ar = \{Ar_1,...,Ar_k,...,
Ar_K\}$ contains the arrival events of any function request of the class $k$, a
set $D=\{D_1,...,D_k,...,D_{K}\}$, and a subset
$D_k=\{D_1^{n},...,D_k^{n},...,D_K^{n}\}$ collects the set of departure events
of a function of class $k$, and $D_k^{n}$ defines the departure event of a
function of class $k$ from node $E_{n}$. 

The resource allocation has the capacity constraints, i.e.,
\begin{equation}\label{CapConst}
\forall n\in [1,N]: \sum_{k=1}^{K}b_k \delta_{k}(n) \leq C_n, 
\end{equation}
The dispatcher (scaler) has three possibilities of actions $a$ to take for every
new event (arrival or departure): to scale up, to scale down or to do nothing.
The action space $A(s)$ is described as follows:
\begin{equation}
A(s)= \begin{cases}
      \{0, 1 \},\;&  e\in Ar \\
     \{-1, 0 \},\; & e\in D \\
    \end{cases}
\end{equation}
where $a(s)=1,\forall k\in \{1,...,K\}$ when a function  of class $k$ is
replicated, $a(s)=-1,\forall k\in \{1,...,K\}$ when a function of class $k$ is
removed from the system, $a(s)=0$ denotes the action of queuing a function
request of any class $k$, without replication in the case of a function arrival,
and the queue update without removing function replica in the case of function
request departure.

\subsection{Transition Probabilities}

At time slot $t_{i+1}$, the $f_k$ request queue length can be expressed as:
\begin{equation}\label{eq:queue}
\begin{split}
Q_k(t_{i+1})= Q_k(t_{i}) + \mathds{1}(a(s)=0 || e=A_k)\\ - \mathds{1}(a(s)=0 || e\in D_k)
\end{split}
\end{equation}
where $Q_k(t_{i})$ describes the previous $f_k$ request queue length when the
$i^\text{th}$ event occurs, $\mathds{1}(.)$ describes the identity operator. The
number of functions of class $k$ replicated in all edge nodes can be expressed
as follows:
\begin{equation}\label{eq:repl}
\begin{split}
\delta_{k}(t_{i+1})= \delta_{k}(t_{i}) + \mathds{1}(a(s)=1 || e=A_k)\\ - \mathds{1}(a(s)=-1 || e\in  D_k^n)
\end{split}
\end{equation}
We assume that the time period between two continuous decision epochs follows an
exponential distribution and denoted as $\tau (s,a)$, given the current state
$s$ and action $a$. Thus the mean rate of events for a specific state $s$ and
action $a$ denoted as $\gamma(s,a)$, is the sum of the rates of all events in
the system, which is expressed as follows:
\begin{equation}
\begin{split}
&  \forall n \in \{1,...,N\}, \forall k \in \{1,...,K\}:\\ & \text{    } \tau(s,a)=  \gamma(s,a)^{-1}= \\  &
 \begin{cases}
       \Lambda + \Theta  + \mu_k  ,\;& e= Ar_k, a=1, \\
     \Lambda +  \Theta , & e= Ar_k, a=0, \\
         \Lambda + \Theta - \mu_k,  & e=D_k^{n}, a=-1,\\
         \Lambda + \Theta,  & e=D_k^{n}, a=0
    \end{cases}
\end{split}\label{eq:tau}
\end{equation}
where $\Lambda = \sum_{k=1}^{K}\lambda_k$ is the total arrival rate of function
requests of all $K$ function classes. When a function request arrives and the
dispatcher decides not to replicate, or a function request leaves an edge node
and the dispatcher decides not to remove a function from the nodes, the total
number of existing functions in the system is $\sum_{k=1}^{K}\delta_{k}$, so the
departure rate of a function in the edge computing is
$\Theta=\sum_{k=1}^{K}\delta_{k}\mu_k$.  When a function request of class $k$ is
replicated, one function of class $k$ is added to the system, thus the departure
rate becomes $\Theta + \mu_k$.  When a departure of a function of class $k$
occurs and the dispatcher removes a replica from the system, the departure rate
becomes  $\Theta - \mu_k$. 

The transition probability in our markov decision model from state $s$ to state
$s'$ when an action $a$ is selected is denotes as $p(s'|s,a)$, which can be
determined under different events:

\paragraph*{i) State $s=(\Delta, Q, Ar_k) $, and $a=0$}
This state describes the system in terms of number of functions allocated in all
nodes, the number of function requests in the queue, and the next event, which
is in this case a function request arrival of class $k$. The function arrival
event can have two types of actions replicate or not to replicate. The following
equation shows the transition probability when the function is queued without
replication, where the number of functions allocated in all nodes remains the
same, and function request queue length $\widehat{Q}$ is updated using eq.
(\ref{eq:queue}), and possible events can occur in the future.  
\begin{equation}
\begin{split}
&p(s'|s,a)= 
 \begin{cases}
     \frac{\lambda_{k'}}{\tau(s,a)}   &   s'=(\Delta, \widehat{Q}, Ar_{k'}) \\
      \frac{\delta_{k'}\mu_{k'}}{\tau(s,a)}   &  s'=(\Delta, \widehat{Q}, D_{k'}^{n}) 
    \end{cases}\label{eq:tr1}
\end{split}
\end{equation}

\paragraph*{ii) State $s=(\Delta, Q, Ar_k) $, $a=1$} This state is defined
similar to the previous state, considering creating a new replica. 
\begin{equation}
\begin{split}
&p(s'|s,a)=
 \begin{cases}
    \frac{\lambda_{k'} }{ \tau(s,a)}  &  s'=(\widehat{\Delta}, Q, Ar_{k'}), \\
    \frac{\delta_{k'}\mu_{k'}}{\tau(s,a)}  &  k'\neq k,  \\ & s'=(\widehat{\Delta}, Q, D_{k'}^{n})  \\
    \frac{(\delta_{k} + 1)\mu_{k}}{\tau(s,a)}  &  s'=(\widehat{\Delta}, Q, D_{k}^{n}) \\
    \end{cases}\label{eq:tr2}
\end{split}
\end{equation}
where $\widehat{\Delta}= \{\delta_{1},...,\delta_{k}+1,...,\delta_{K} \}$.

\paragraph*{iii) State $s=(\Delta, Q, D_k^n) $, and $a=0$} 
 The next event in this state is a function request departure of class $k$ from
edge node $E_{n}$. The function departure event can have two types of actions
remove or not remove a function replica. The following equation shows the
transition probability when the function leaves the system without removing a
function replica, where the number of functions allocated in all edge nodes
remains the same, and function request queue length $\widehat{Q}$ is updated
using eq. (\ref{eq:queue}), and possible events can occur in the future.  
\begin{equation}
\begin{split}
&p(s'|s,a)= 
 \begin{cases}
     \frac{\lambda_{k'}}{\tau(s,a)}   &   s'=(\Delta, \widehat{Q}, Ar_{k'}) \\
      \frac{\delta_{k'}\mu_{k'}}{\tau(s,a)}   &  s'=(\Delta, \widehat{Q}, D_{k'}^{n}) 
    \end{cases}\label{eq:tr3}
\end{split}
\end{equation}

\paragraph*{iv) State $s=(\Delta', Q, D_k^n) $, $a=-1$} This state is defined
similar to the previous state, considering removing a replica.

\begin{equation}
\begin{split}
&p(s'|s,a)=
 \begin{cases}
    \frac{\lambda_{k'} }{ \tau(s,a)}  &  s'=(\widehat{\Delta}, Q, Ar_{k'}), \\
    \frac{\delta_{k'}\mu_{k'}}{ \tau(s,a)}  &  k'\neq k\\ & s'=(\widehat{\Delta}, Q, D_{k'}^{n})  \\
    \frac{(\delta_{k} - 1)\mu_{k}}{ \tau(s,a)}  &  s'=(\widehat{\Delta}, Q, D_{k}^{n}) \\
    \end{cases}\label{eq:tr4}
\end{split}
\end{equation}
where  $\widehat{\Delta}= \{\delta_{1},...,\delta_{k}-1,...,\delta_{K} \}$.

\subsection{Rewards}

Given the system state $s$ and the corresponding action $a$,  the system reward
of the function provider is denoted by
\begin{equation}\label{eq:reward}
r(s, a)=w(s,a) - g(s,a)
\end{equation}
where $w(s,a)$ is the net lump sum incomes of the system at the  state $s$ when
action $a$ is taken and an event $e$ occurs, and $g(s,a)$ is the expected system
costs.
\begin{equation}
\begin{split}
w(s,a)= 
 \begin{cases}
       w_k  & e= Ar_k, a\in \{0, 1\}\\
       0  & e=D_k^{n}, a=\{0, -1\} \\
    \end{cases}
\end{split}
\end{equation}
where the variable $w_k$ denotes the reward of the  system for accepting of a
function request of class $k$. The expected system cost $g(s,a)$ is defined as: 
\begin{equation}\label{eq:g}
g(s,a)= c(s, a)\cdot \tau (s, a)
\end{equation}
where $\tau(s,a)$ is the expected service time defined by eq. (\ref{eq:tau})
from the  state $s$ to the next state in case that action $a$ is chosen and
$c(s,a)$ is the service holding cost rate when the system is in state $s$ in
case that action a is selected, which depends on  the queuing. Furthermore,
$c(s,a)$ can be described by the number of occupied resources in the system, the
queue length using Little's Law, as follows: 
\begin{equation}\begin{split}
&c(s,a)= \underbrace{\sum_{k=1}^{K}c\cdot b_{k}\cdot \delta_{k}}_\text{Processing} + \underbrace{\sum_{k=1}^{K} \frac{Q_k}{\lambda_k}}_\text{Queuing} 
\end{split}
\end{equation}
where $c$ represents the utilization cost of a resource unit. In order to only
optimize the delay, the processing cost can be ignored by setting $c$ to $0$.

\subsection{SMDP-based Scaling Model}\label{subsec:SMDP} 

We develop an SMDP-based Scaling Model (SM) to study the performance of a serverless
platform considering the queuing and processing delay of function requests. We
aim at taking the optimal decisions at every decision event (arrival of new function
request, and departure of a function request) where our goal is to maximize the
long-term expected system rewards. The expected discounted reward is given based
on the model in \cite{puterman2014markov} as follows:
\begin{equation}\label{eq:discount}
\begin{split}
r(s,a)=&w(s,a)-c(s,a)\cdot E_{s}^{a}\lbrace \int_{0}^{\tau} e^{-\alpha t} dt\rbrace\\
=&w(s,a)-c(s,a)\cdot E_{s}^{a}\lbrace \frac{1-e^{-\alpha \tau}}{\alpha}\rbrace\\
=&w(s,a)- \frac{c(s,a) }{\alpha + \tau(s,a)}\\
\end{split}
\end{equation}
where $\alpha$ is a continuous-time discount factor.

Using the defined transition probabilities eq. (\ref{eq:tr1}), (\ref{eq:tr2}),
(\ref{eq:tr3}), (\ref{eq:tr4}), we can obtain the maximum long-term discounted
reward using a discounted reward model defined in \cite{puterman2014markov} as
\begin{equation}\label{eq:discount2}
\nu(s)=\max_{a\in A(s)}\left\lbrace r(s,a) + \lambda \sum_{s'\in S} p(s'|s,a) \nu(s') \right\rbrace
\end{equation}
where $\lambda=\tau(s,a) / (\alpha+\tau(s,a)) $. In the SMDP model, the value of
$\nu (s)$ in a strategy $\psi$ is computed based on the value $\nu (s')$
obtained in the strategy $\psi - 1$, and as an initial value, the discounted
reward can be set to zero for all states to initialize the computation, which
converges afterwards towards the optimal solution.

To simplify the computation of the reward, let $\rho$ be a finite constant,
where $\rho= \sum_{i=1}^{K}\lambda_i   +  \sum_{n=1}^{N}\sum_{k=1}^{K}C_n \mu_k
< \infty$. We define $\overline{p}(s'|s,a)$, $ \overline{\nu}(s)$, and
$\overline{r}(s,a)$  as the uniformed transition probability, long-term reward,
and reward function, respectively, and given by:
\begin{equation}
\overline{r}(s,a)= r(s,a)\frac{\tau (s,a) + \alpha}{\rho+\alpha}, \overline{\lambda}= \frac{\rho}{\rho+\alpha}
\end{equation}
\begin{equation}
\begin{split}
\overline{p}(s'|s,a)= 
 \begin{cases}
       1-\frac{[1-p(s'|s,a)] \tau (s,a)}{\rho}  & s'=s\\
       \frac{p(s'|s,a) \tau (s,a)}{\rho} & s'\neq s\\
    \end{cases}
\end{split}
\end{equation}
After uniformization, the optimal reward is given by:
\begin{equation}\label{eq:optreward}
\overline{\nu}(s)=\max_{a\in A(s)}\left\lbrace \overline{r}(s,a) + \overline{\lambda} \sum_{s'\in S} \overline{p}(s'|s,a) \overline{\nu}(s') \right\rbrace
\end{equation}
In order to solve our SMDP-SM, we use the
iterative algorithm described as follows:  

\begin{algorithm}
\caption{Iterative SMDP-SM Algorithm}
\label{alg:SMDPSPalgorithm}
\begin{algorithmic}[1]
\State \textbf{Step 1 (Initialization):} $\overline{\nu}^0(s)=0$, for all $s\in S$. Set the value of  $\epsilon >0$, and iteration $t=0$.
\State \textbf{Step 2:} Using eq. \ref{eq:optreward}, compute the discounted reward for each state $s$:
$$\overline{\nu}^{t+1}(s)=\max_{a\in A(s)}\left\lbrace \overline{r}(s,a) + \overline{\lambda} \sum_{s'\in S} \overline{p}(s'|s,a) \overline{\nu}^{t}(s') \right\rbrace$$
\State \textbf{Step 3:}\If{$\| \overline{\nu}^{t+1} - \overline{\nu}^{t} \|> \epsilon$ } $t\longleftarrow t+1$,  go to \textbf{Step 2}
\Else  \; go to \textbf{Step 4} 
\EndIf
\State \textbf{Step 4:} Compute the optimal scaling policy for all $s\in S$  
$$d_{\epsilon}^{*}(s)\in \text{arg}\max_{a\in A(s)}\left\lbrace \overline{r}(s,a) + \overline{\lambda} \sum_{s'\in S} \overline{p}(s'|s,a) \overline{\nu}^{t+1}(s') \right\rbrace$$
\end{algorithmic}
\end{algorithm}

After obtaining the optimal policy from Algorithm \ref{alg:SMDPSPalgorithm}, the
steady states probabilities are computed as, i.e.,
\begin{equation}
\begin{split}
\pi (P-J)=0,  \sum_{s\in S}\pi(s)=1
\end{split}
\end{equation}
where $\pi(s)$ represents the steady state probability at state $s$, $P$ is the
transition probabilities matrix, considering the optimal policy
$d_{\epsilon}^{*}$, and $J$ denotes the all-ones matrix.

The \textbf{Algorithm} \textbf{\ref{alg:SMDPSPalgorithm}} is executed offline after defining the network topology in order to find the optimal scaling policies for every possible system state. Afterward the optimal policies are used online as a look-up table to make scaling decision with a time complexity of $O(1)$.

\subsection{Complexity Analysis}

Consider the case with $N$ nodes, $K$ functions, the time complexity of
\textbf{Algorithm}  \textbf{\ref{alg:SMDPSPalgorithm}}, which is based on Policy
Iteration algorithm depends on the number of states $|S|$, number of actions
$|A|$ and the discount factor $\gamma$. Scherrer   \cite{scherrer2013improved}
has proven that Policy Iteration terminates after at most $O \left( \frac{|S|
\cdot |A|}{1-\gamma} \log\left( \frac{1}{1-\gamma}  \right) \right) $.  The
number of states can be represented in terms of the network configuration
parameters using eq. (\ref{eq:state}), as combination of possible number of
replicas of each function, the state of the queues, and the number of possible
next events (arrival or departure). Assuming that the system can have a maximum
number of replicas per function denoted as $M$, and a maximum queue length
$Q_m$,  $|S|=M^K \cdot Q_m^K \cdot  (K + K\cdot N)$. The number of actions in
our proposal is equal to 3. Thus the time complexity of
\textbf{Algorithm}.\ref{alg:SMDPSPalgorithm} is given by:
\begin{equation}\label{expr:time_complexity}
O \left( \frac{M^K \cdot Q_m^K \cdot  K\cdot (N + 1) }{1-\gamma} \log\left( \frac{1}{1-\gamma}  \right) \right)
\end{equation}
Space complexity of the algorithm is mainly driven by the storage of the states
information generated at the initialization phase and given by:
\begin{equation}\label{expr:space_complexity}
O \left( M^K \cdot Q_m^K \cdot  K\cdot (N + 1)  \right)
\end{equation}

\section{Numerical results}\label{sec:results}

We now evaluate the performance of the SMDP scaling methods, by comparing it to
the monitoring-based heuristics, and for the sake of verification, by comparing
it to a random-fit model. The monitoring-based methods are relevant to
evaluation, since they are used today, e.g., in OpenFaaS. In OpenFaaS, for
instance, the scaling decisions consider arrival rates by default. In practice,
the theoretical time complexity is in fact $O(1)$. As shown in Fig. \ref{fig:system}, the serverless platform includes the monitoring-based scaling, where the telemetry collects information about the load.  Whenever the load exceeds a certain threshold, the monitoring algorithm triggers the system to create new replicas, and whenever the queue becomes empty, the algorithm decides to remove replicas, otherwise it keeps the same system state. The monitoring-based
algorithm is thus described as follows:
 
\begin{algorithm}
\caption{Monitoring-based scaling algorithm}
\label{alg:MBalgorithm}
\begin{algorithmic}[1]
\State \textbf{Input: } event, queue, capacity, load, threshold
\If{event is an arrival }  \If{capacity is available}  \If{load > threshold}  return 1 
\Else  \;return 0
\EndIf 
\Else  \;return 0
\EndIf
\EndIf
\If{event is a departure }  \If{the queue is empty }   return -1
\Else  \; return 0
\EndIf
\EndIf
\end{algorithmic}
\end{algorithm}

For comparison purposes, we also show random-fit scaling (theoretical time
complexity $O(1)$), that decides randomly either to scale or not, if there are
available resources. The algorithm follows the same structure as in the
monitoring based approach with the only difference that in line 4, instead of
checking the threshold, we just return a random node. 


We evaluate the performance of the three scaling methods using: i) \emph{SMDP}
model, ii) monitoring-based (\emph{MNT}) and iii) random-fit (\emph{RF}). Since
scaling decisions do not determine function allocations, we assume two simple
allocation approaches: First-Fit allocation (FFa) and Random-Fit allocation
(RFa). In FFa, functions are allocated in the closest available node,
considering a network with a single master and multiple workers. In RFa,
functions are allocated randomly in any available node with enough available
resources.  All three scaling methods are then evaluated considering both FFa
and RFa. We evaluate two networks, a small one with 3 edge nodes and a big one
with 10 nodes. We use the small one for comparing all methods, with threshold
numbers $0.1$ and $0.05$. And, we use the large network to compare the
performance of only heuristics, with threshold numbers $0.01$, $0.005$ and
$0.001$. For generating the results, we use an event based simulator running up
to 1 million events with severals seeds to validate our results. The rest of the
parameters are shown in Table \ref{tab:param}. 

\begin{table}[!ht]
  \begin{center}
    \caption{Simulation Parameters.}
    \label{tab:param}
    \begin{tabular}{lcccccccc} 
      \toprule
    \textbf{Parameters}  & $N$  &  $K$  & $b_k$ &$C_{n}$ & $c$  & $w_k$  &  $\lambda_k$  & $\mu_k$ \\
    \midrule
    \textbf{Small network}  & 3  &  5  &  $k$ & 16 & 1  & 1  &  2-11  & 1-11 \\
    \textbf{Large network}  & 10  &  10  &  $k$ & 100 & 1  & 1  &  4-12  & 10-$10^2$ \\
      \bottomrule
    \end{tabular}
  \end{center}
\end{table}

\begin{figure}[!ht]
  \centering
    \includegraphics[width=0.475\textwidth]{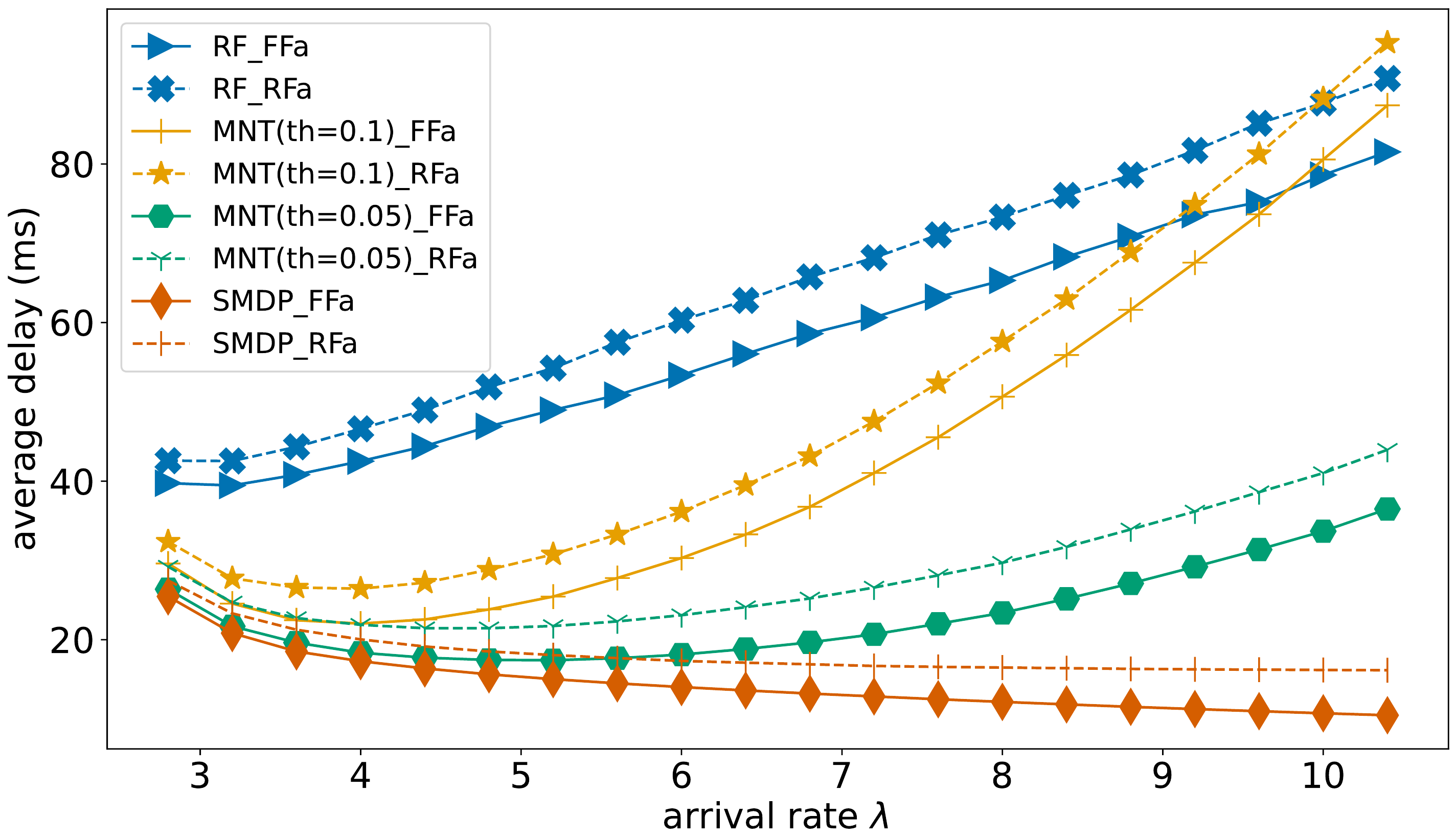}
  \caption{Average service delay per scaling model (small network).}
 \label{fig:avgr-service-delay_small}
 \end{figure}

Fig. \ref{fig:avgr-service-delay_small} shows the average service delay of all
functions for arrival rates of requests for each scaling algorithm in the
small network. The results shows that \emph{SMDP} overperforms all other
approaches for any value of arrival rate. The delay obtained by monitoring based
approach decreases with the decrease of the threshold value and outperforms the
random approach until a certain value of arrival rate $\lambda$, after which the
delay starts to increase exponentially, at $\lambda=9.6$ in Figure
\ref{fig:avgr-service-delay_small}. This behavior is created by the accumulation
of function requests in the queue when the load becomes constantly lower than
the threshold, resulting a cumulative queueing delay. The delay results using
First-Fit allocation is slightly better than the results obtained with a
Random-Fit allocation, which can be explained by the fact that the average delay
is mostly affected by the queueing and less by the transmission delay.  

\begin{figure}[!ht]
  \centering
    \includegraphics[width=0.475\textwidth]{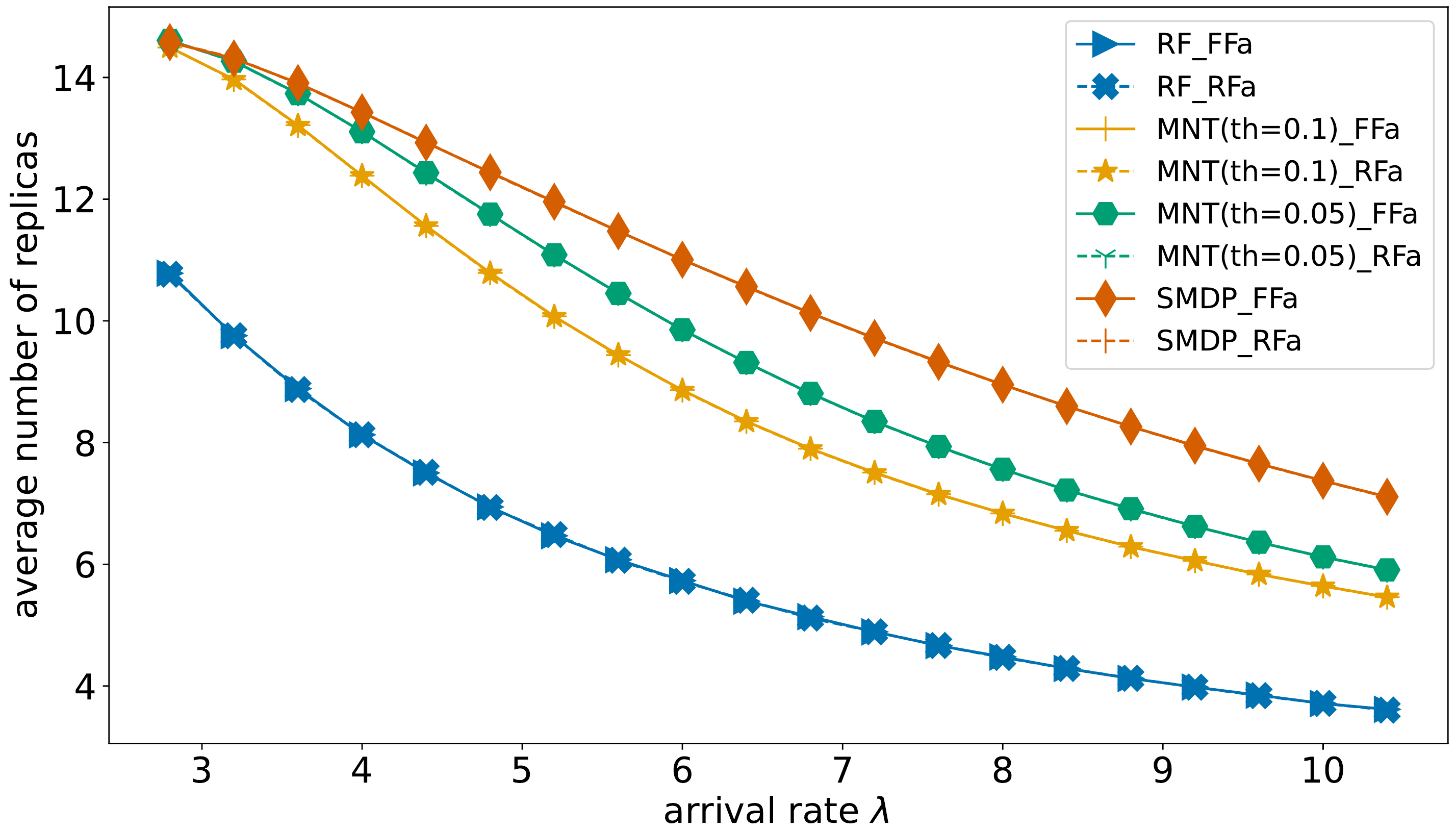}
  \caption{Average number of replicas per scaling model (small network).}
 \label{fig:avgr-nb-rep_small}
 \end{figure}

Figure \ref{fig:avgr-nb-rep_small} shows the average number of replicas of all
functions with different arrival rates of function requests for each scaling
and allocation algorithm. The results shows that the chosen algorithm does not
affect the number of replicas of each function, where both curves coincide. The
highest average number of replicas is obtained with SMDP, which explains its
performance in terms of  delay. As expected, the lowest number of replicas is
obtained by the \emph{RF} approach.

\begin{figure}[!ht]
  \centering
    \includegraphics[width=0.49\textwidth]{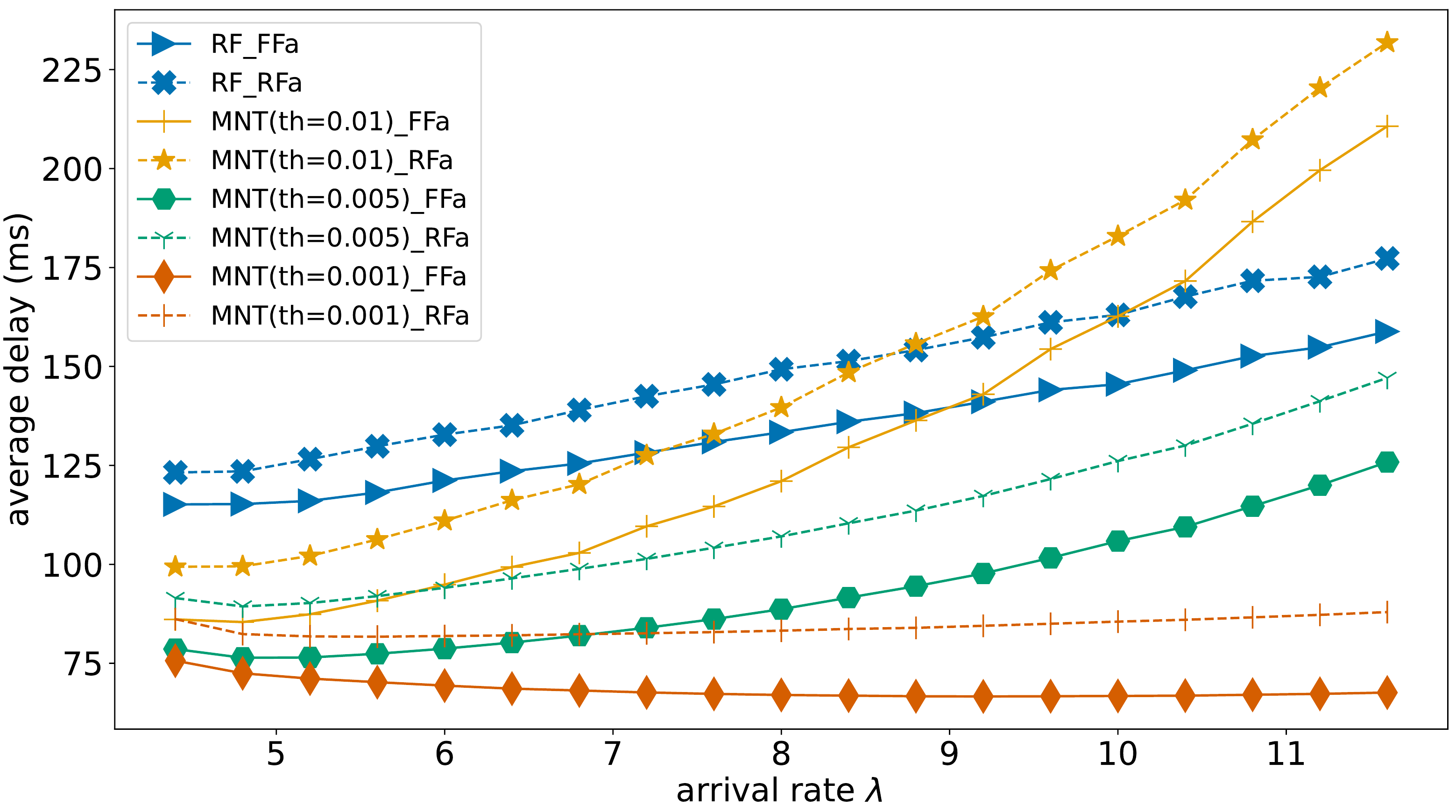}
  \caption{Average service delay per scaling model (large network).}
 \label{fig:avgr-service-delay_big}
 \end{figure}

\begin{figure}[!ht]
  \centering
    \includegraphics[width=0.49\textwidth]{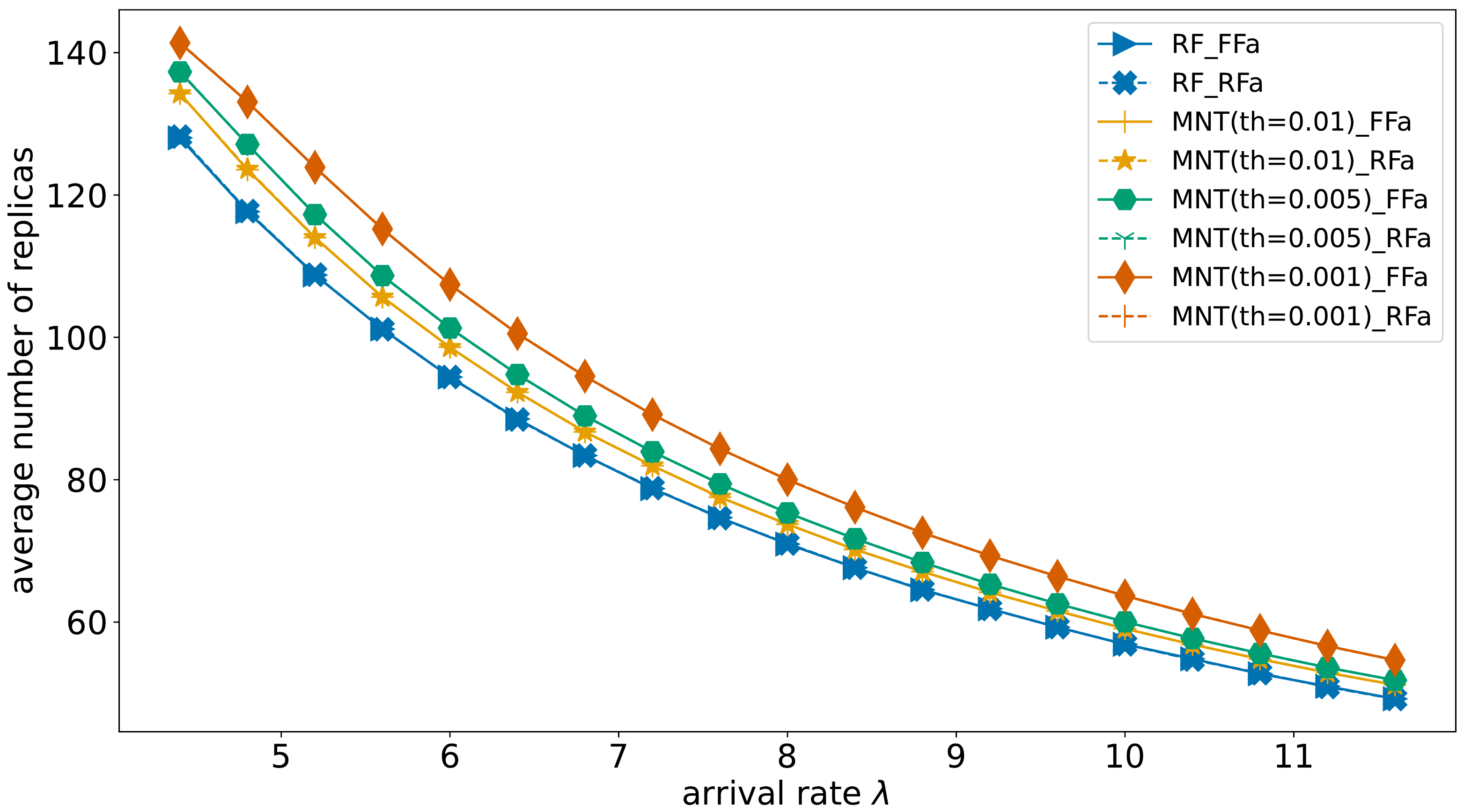}
  \caption{Average number of replicas per scaling model (large network).}
 \label{fig:avgr-nb-rep_big}
 \end{figure}

Figure \ref{fig:avgr-service-delay_big} illustrates the average service delay of
all functions with different arrival rates of function requests. The
monitoring-based approach with a very low threshold value $0.001$ outperforms
the random approach, where its performance improves while decreasing the
threshold value. A similar behavior is found in the small network for high
threshold values, where monitoring-based approach outperforms the random
approach  until a certain value of arrival rate $\lambda$, after which the delay
starts to increase exponentially. Finally, we illustrate in Figure
\ref{fig:avgr-nb-rep_big} the average number of replicas for the large network.
Similar to the small network, the allocation algorithm choice does not affect
the number of replicas and Monitoring-based approach with the lowest threshold
value creates larger number of replicas.

 \section{Conclusion}\label{sec:conclusion}
 
In this paper, we addressed the optimality of serverless scaling in edge computing network and proposed to use  Semi-Markov Decision Process-based (SMDP) model for the scaling problem of serverless functions as a decision making problem, with actions of scaling the functions up or
down. The theoretical results were compared with practical, monitoring-based algorithms based on current approaches.The results confirmed that SMDP gave best results in terms of queuing delay, and outperformed
monitoring-based approaches. The monitoring-based
approach however achieved performance comparable to the optimal SMDP solution in terms of
delay when the scaling activation threshold was set to comparably lower values.
 In our future work, we will study the joint scaling and resource allocation problem
considering various system parameters. SMDP can be used to analyse the optimal
policy for a specific setting. Monitoring based is what is used in real
implementations now, which can be more optimized using smarter algorithms such
as SMDP. The issue of SMDP is its exponential expansion, so a heuristic based on
SMDP can be a very good approach for
future.

%

\section*{Acknowledgment}
This work was partially supported by EU HORIZON research and
innovation program, project ICOS (Towards a functional continuum operating system), Grant Nr. 101070177.
\bibliographystyle{IEEEtran}
\bibliography{mybib}

\end{document}